\documentclass[aps,pre,twocolumn,showpacs]{revtex4-1}
\usepackage{amsmath}
\usepackage{epsfig}
\usepackage{amssymb}
\usepackage{dcolumn}
\usepackage{graphicx}
\usepackage{dcolumn}
\usepackage{graphics}
\usepackage{color}

\begin{document}

\title{Modulational instability windows in the nonlinear Schr\" odinger equation involving higher-order Kerr responses}

\author{David Novoa$^1$, Daniele Tommasini$^{2}$, and Jos\'e A. N\'ovoa-L\'opez$^2$}
\affiliation{ $^1$Max Planck Institute for the Science of Light, G\"unther-Scharowsky str. 1, 91058 Erlangen, Germany;\\
$^2$Departamento de F\'{i}sica Aplicada. Universidade de Vigo. As Lagoas s/n, 32004 Ourense, Spain.}

\begin{abstract}

We introduce a complete analytical and numerical study of the modulational instability process in a system governed by a canonical nonlinear Schr\"odinger equation involving local, arbitrary nonlinear responses to the applied field. In particular, our theory accounts for the recently proposed higher-order Kerr nonlinearities, providing very simple analytical criteria for the identification of multiple regimes of stability and instability of plane-wave solutions in such systems. Moreover, we discuss a new parametric regime in the higher-order Kerr response which allows for the observation of several, alternating stability-instability windows defining a yet unexplored instability landscape.

\end{abstract}


\maketitle

\section{Introduction}
Modulational instability (MI) is a nonlinear phenomenon causing a plane wave (PW) or long pulse to break up into smaller, finite sub-structures. It can happen in a plethora of different physical systems, ranging from deep water waves~\cite{BenjaminFeir66} to optical beams~\cite{BT}. In particular, the mechanism of MI in optics, which may happen in both temporal~\cite{MIrmp} and spatial domains~\cite{BT,AgrawalKivshar}, either independently or simultaneously, has been extensively studied due to its paramount implications on the dynamics of intense light beams or pulses traveling throughout nonlinear media. In brief, perturbations in the incoming optical field can be amplified upon propagation due to the nonlinear response of the optical medium~\cite{BT}, leading to the reorganization of the energy and subsequent generation of finite optical beams or \emph{filaments}, randomly distributed along the space-time profile of the parent wave.

Regarding their effects, temporal MI of either continuous waves or very long pulses is responsible for the generation of ultrashort pulses~\cite{MIrmp} with temporal durations down to- or even below the single-cycle limit~\cite{extremeMI}. Such MI-driven short pulses are linked to the appearance of spectral sidebands which, in combination with other nonlinear processes, can boost the generation of extremely broadband radiation. In contrast, spatial MI has been demonstrated to be, in general, an unwanted effect in the propagation of intense ultrashort pulses throughout bulk optical media. In fact, the MI-triggered redistribution of the energy along the transverse profile of the beams gives rise to the appearance of intense hot spots that could potentially damage the optical material itself (see e.g.~\cite{couairon05} and references therein). Each intense spot contains an amount of power comparable to the critical power $P_{cr}$ for self-focusing~\cite{selffocusing}. Whilst such structures are doomed to undergo collapse due to self-focusing whenever their power $P\geq P_{cr}$ in a Kerr medium, they have been shown to form stable multidimensional solitary waves when traveling across optical media with competing cubic and quintic (CQ) nonlinearities~\cite{filamentation}, yielding in some cases to the formation of \emph{liquid light} states~\cite{michinel02}, i.e., intense flat-top beams featuring intriguing surface tension properties~\cite{novoa09}. Very remarkably, this new type of self-trapped beams have recently been observed experimentally in coherently-engineered atomic media
~\cite{llightexp}.

On the other hand, the recent measurement of Higher-Order Kerr (HOKE) responses in air and its constituents~\cite{HOKEexp} gave rise to an extensive, ongoing discussion on the physical origin and validity of such peculiar contribution to the nonlinear polarization~\cite{kolesik10, bree10, polyinkin11, bejot11, wahlstrand11a, wahlstrand11b, bejot13, kohler13}. Among other reasons, such a debate has been specially motivated by the deep implications of the HOKE response in the description of laser filamentation~\cite{HOKEteor} and novel light distributions~\cite{fermionic}. In particular, theoretical media displaying such kind of competing HOKE terms have been shown to allow for a new branch of highly concentrated nonlinear solitary waves, called \emph{ultrasolitons}~\cite{ultrasolitons}, which might coexist with the usual solitons appearing in CQ-like nonlinear models~\cite{CQsolitons}.  

In this paper, we introduce a complete analytical and numerical study of the MI process in a system governed by a canonical nonlinear Schr\"odinger equation (NLSE) involving a local, arbitrary nonlinear response $F$ to the applied field. In particular, in Sec. II we will revisit the theory introduced in~\cite{BT} to account for the recently proposed HOKE nonlinearities, deriving very simple analytical criteria for the identification of multiple regimes of stability and instability of PW solutions in such systems. To illustrate our results, in Sec. III we will show that the new model explains the phenomenology reported in the literature for different kinds of nonlinear responses. In addition, we will discuss the parametric regime of HOKE responses proposed in~\cite{ultrasolitons}, which allows for the existence of ultrasolitons. In this case, we will show that the theory predicts several alternating stability-instability windows whose existence 
is also demonstrated by means of numerical simulations in Sec. IV. Thus, our results define a completely new MI landscape as compared to the scenario that has recently been described~\cite{generalMI} in the presence of HOKE nonlinearities out of the multistability regime.

\section{General theory}

\subsection{The canonical scalar NLSE and its plane wave solutions}

We consider a system described by the wave function $\Psi({\bf r},\eta)$ evolving along the $\eta$ direction. Such a system can be either a nonlinear optical medium in which $\Psi({\bf r},\eta=z)$ would represent the scalar electric field envelope of an optical wave propagating along the $z$ direction~\cite{AgrawalKivshar}, or an ultracold atomic gas in which $\Psi({\bf r},\eta=t)$ describes the order parameter of the corresponding macroscopic collective quantum state evolving in time $t$~\cite{BEC}. In such a systems, the evolution of $\Psi({\bf r},\eta)$ in a $(n+1)$-dimensional space of points $({\bf r},\eta)$ is governed by the canonical nonlinear Schr\"odinger equation

\begin{equation}
  \label{NLSE}
i \frac{\partial \Psi}{\partial \eta} + \frac{1}{2}
  \nabla^2 \Psi + F(\vert\Psi\vert) \Psi = 0,
\end{equation}
where $\nabla^2 =\partial^2 / \partial {\bf r}^2$ accounts for diffraction (dispersion) in the spatial (temporal) domain and $F$ is an arbitrary, real-valued and continuous function of $\vert\Psi\vert$. 
For the sake of simplicity, we will restrict our discussion to Eq. \eqref{NLSE}, although our approach can be easily generalized in the presence of terms involving higher order derivatives, such as those that should be introduced to describe the propagation of ultrashort pulses~\cite{generalMI}. 

We assume that all the quantities in Eq. \eqref{NLSE}, including the coordinates ${\bf r}$, have been suitably rescaled in such a way that they are dimensionless. All the relevant parameters will then appear in the explicit form of the function $F$.

It is well-known that Eq. \eqref{NLSE} admits PW solutions of the form 
\begin{equation}
  \label{plane-wave} 
\Psi=A \exp(-i \mu \eta+i\varphi_0),
\end{equation}
where $A>0$, $\mu$ and $\varphi_0$ are real constants describing the amplitude, propagation constant (chemical potential in case of matter-waves) and global phase, respectively. 
Eq. \eqref{NLSE} implies the relation
\begin{equation}
  \label{mu-FA}
\mu=-F(A),
\end{equation}
which reflects the nonlinear behavior of the system since the phase of the wave is \emph{self-modulated} by its intensity~\cite{AgrawalKivshar}. In addition, due to the U(1) symmetry of Eq. \eqref{NLSE}, its different solutions are invariant under global phase transformations and so the arbitrary phase factor $\varphi_0$ will play no role in the following discussion.

In the next sections, we will analyze the dynamics of PWs propagating through optical media governed by Eq. \eqref{NLSE} in the presence of noise. In particular, we will derive a fairly simple analytical rule for the prediction of the onset of MI. Such a rule is completely general, and can be applied to a plethora of nonlinear systems described by different nonlinear responses $F$.

\subsection{Modulational instability and perturbation growth rate}

Let us now study the stability of the PW solutions of Eq. \eqref{NLSE} under the influence of small perturbations by generalizing the perturbative method of Ref.~\cite{BT}. Similar analyses have also been carried out in Refs.  \cite{MI1,MI2,MI3,MI4,MI5}.
We will then look for a solution of the form $ \Psi({\bf r},\eta)=\left(A+f({\bf r},\eta)\right) \exp(-i \mu \eta+i\varphi_0)$, where $f({\bf r},\eta)$ is an arbitrary, complex-valued perturbation. An additional interesting possibility, which lies beyond the scope of the present paper, would correspond to the consideration of localized perturbations on finite backgrounds \cite{pert1,pert2}.

The requirement that the pure PW solution is stable under the action of small perturbations corresponds to the condition that $\vert f({\bf r},\eta)\vert$ can be kept $\vert f({\bf r},\eta)\vert\ll A$ for all values of ${\bf r}$ and $\eta$. In this case, at the first order in $f$,  
$$
\vert\Psi\vert^2=\vert A+f\vert^2  = A^2 + 2 A \alpha + \alpha^2 + \beta^2 \simeq A^2(1 + 2 \alpha/A),
$$
where $\alpha$ and $\beta$ are real functions describing the real and imaginary parts of $f$, such that $f({\bf r},\eta)=\alpha({\bf r},\eta)+i\beta({\bf r},\eta)$, and $\vert\alpha\vert, \vert\beta\vert \ll A$.
Therefore $\vert\Psi\vert\simeq A+ \alpha$, and Eq. \eqref{NLSE} becomes
\begin{equation}
  \label{eq-stability-cond}
i \frac{\partial f}{\partial \eta} + \frac{1}{2}
  \nabla^2 f + \alpha A F'(A) = 0,
\end{equation}
where $F'(A) =\frac{\partial F}{\partial \vert\Psi \vert}(A) $ and we have taken into account Eq. \eqref{mu-FA}. By separating the real and imaginary parts, we get a set of two coupled equations,
\begin{eqnarray}
  \label{eq-stability-cond-re+im} 
&\frac{\partial \alpha}{\partial \eta} + \frac{1}{2}
  \nabla^2 \beta=0,\cr
&
- \frac{\partial \beta}{\partial \eta} + \frac{1}{2}
  \nabla^2 \alpha+  A F'(A) \alpha = 0.
\end{eqnarray}

This system of equations can be solved in the reciprocal space by expanding $\alpha({\bf r},\eta)=\int_{-\infty}^{\infty} \tilde\alpha({\bf k},\eta)\exp(i {\bf k}\cdot{\bf r})d^{n} {\bf k} $ and $\beta({\bf r},\eta)=\int_{-\infty}^{\infty} \tilde\beta({\bf k},\eta)\exp(i {\bf k}\cdot{\bf r})d^{n} {\bf k} $, where ($\tilde\alpha$, $\tilde\beta$) stand for the Fourier transforms of ($\alpha$, $\beta$) and $n$ is the dimension of the space of the vectors ${\bf r}$. We then obtain $\tilde\beta=\frac{2}{k^2}\frac{\partial \tilde\alpha}{\partial \eta}$, being $k\equiv\vert {\bf k}\vert$,  together with the following equation

\begin{equation}
  \label{eq-stability-fourier} 
\frac{\partial^2 \tilde\alpha}{\partial \eta^2} +\omega^2(k)\tilde\alpha = 0,
\end{equation}
which is formally similar to the equation governing the dynamics of a classical harmonic oscillator whose angular frequency $\omega(k)$ satisfies the following dispersion relation
\begin{equation}
  \label{omegak} 
\omega^2(k)\equiv\frac{k^2}{2}\left[\frac{k^2}{2}-  A F'(A) \right].
\end{equation}

Whenever $\omega(k)$ becomes pure imaginary for a certain value of $k$, the perturbations described by ($\alpha$, $\beta$) experience an exponential amplification. In this case, the corresponding PW will be unstable and the perturbative approximation will be no longer fulfilled. On the other hand, if $\omega(k)$ is real for all wavevectors $k$,  the values of $\alpha$ and $\beta$ will be constrained by the initial conditions. In other words, the perturbations $f({\bf r},\eta)$ can be kept small at all points of the space $({\bf r},\eta)$ if and only if $\omega(k)$ is a real number for all values of $k$. As a consequence, the stability condition can be expressed in the following very simple form
\begin{equation}
  \label{stability-cond}
F'(A)<0.
\end{equation}

Accordingly, whenever $F'(A)>0$ the corresponding PW solution will not be stable and will eventually break up into small localized substructures or filaments~\cite{BT}. In the latter situation, it is very useful to compute the wavevector $k_{\rm max}$ related with the fastest growing Fourier component by maximizing the value of the exponential growth factor $\Gamma(k)\equiv i \omega(k)=\frac{k}{\sqrt2}\left[AF'(A)-\frac{k^2}{2}\right]^{1/2}$. We obtain
\begin{equation}
  \label{kmax}
k_{\rm max}=\left[A F'(A)\right]^{1/2}.
\end{equation}
This will be our theoretical prediction for the location of the peaks of the MI sidebands, as they are called in optics. The corresponding prediction for the maximum exponential growth rate, called maximum MI gain in the context of nonlinear optics, will then be

\begin{equation}
  \label{Gammamax} 
\Gamma_{\rm max}\equiv \Gamma(k_{\rm max})=\frac{k_{\rm max}^2}{2}=\frac{AF'(A)}{2}.
\end{equation}

For $F'(A)>0$, and in the presence of noise, the perturbed PW will break into filaments having a characteristic size given by $\Lambda=\pi/k_{\rm max}=\pi/\sqrt{2\Gamma_{max}}$~\cite{BT}. Notice that the expression for $\Lambda$ is only valid at the very first stages of the wave destabilization.


\section{Examples of application. HOKE nonlinearities.}

In this section, we will apply the general formalism derived above to some canonical nonlinear systems like the Kerr and Cubic-Quintic (CQ) nonlinear media~\cite{filamentation}. In particular, we will show that the general expression of Eq. \eqref{stability-cond} reproduces the known results on the MI dynamics for such media. In addition, we will discuss the case of a generalized HOKE nonlinearity, and we will find the appearance of different instability windows, depending on the relative strengths of the higher order nonlinear terms.

\subsection{Cubic nonlinearity}

The cubic nonlinearity (called \emph{Kerr} nonlinearity in optics) naturally appears both in the mean-field description of Bose-Einstein condensates and in the modeling of a large class of optical systems. In this case, 

\begin{equation}
F(\vert \Psi\vert)=f_ 2 \vert \Psi\vert^2,
\end{equation}
and the constant $f_2$ can be chosen to be $\pm1$ by suitably rescaling $\Psi$.  
The stability condition Eq. \eqref{stability-cond}  can then be written as $f_2<0$, corresponding to the self-defocusing Kerr nonlinearity. In the opposite case,  $f_2>0$, the PWs will undergo MI for any value of the amplitude $A$. The fastest growing perturbations leading to the break-up of the plane wavefront in multiple filaments will correspond to a wavevector 

\begin{equation}
k_{\rm max}=\sqrt{2}A,
\end{equation}
and the corresponding value of the maximum perturbation growth rate will accordingly be
\begin{equation}
  \label{Gmax} 
\Gamma_{\rm max}=A^2.
\end{equation}

As expected, these results agree with those obtained in the original paper by Bespalov and Talanov~\cite{BT}.

\subsection{Cubic-Quintic nonlinearity}

CQ materials have been thoroughly studied in the literature from the theoretical point of view~\cite{CQsolitons}, due to the interesting properties they display as a result of their nonlinear response~\cite{michinel02,novoa09}. Remarkably, the first realization of a canonical CQ nonlinearity has recently been accomplished in coherent atomic optical media~\cite{llightexp}.

By suitably rescaling $\Psi$ and the space-time coordinates, light propagating throughout CQ systems can be described by Eq. \eqref{NLSE}  involving a nonlinear term

\begin{equation}
F(\vert \Psi\vert)=f_2(\vert \Psi\vert^2- \vert \Psi\vert^4 )
\end{equation}
where $f_2=\pm1$.

Taking into account that we have chosen $A>0$, the stability condition Eq. \eqref{stability-cond}  can then be written as $f_2 (1-2A^2)<0$, i.e.  $A>1/\sqrt{2}$ for $f_2=+1>0$, or $A<1/\sqrt{2}$ for $f_2=-1<0$. 

Again, the opposite case, $f_2 (1-2A^2)>0$, i.e.  $A<1/\sqrt{2}$ for $f_2=+1$, or $A>1/\sqrt{2}$ for $f_2=-1$, will yield to MI of the incoming PWs, and the fastest growing perturbations will then feature a wavevector

\begin{equation}
k_{\rm max}=\sqrt{2}A\sqrt{\vert1-2A^2\vert},
\end{equation}
being the maximum perturbation growth rate 
\begin{equation}
  \label{Gmax_cubic}
\Gamma_{\rm max}=\vert A^2-2A^4\vert.
\end{equation}

These results agree with those obtained in Ref.~\cite{filamentation}.


\begin{figure}[htbp!]
\centerline{\includegraphics[width=1\columnwidth]{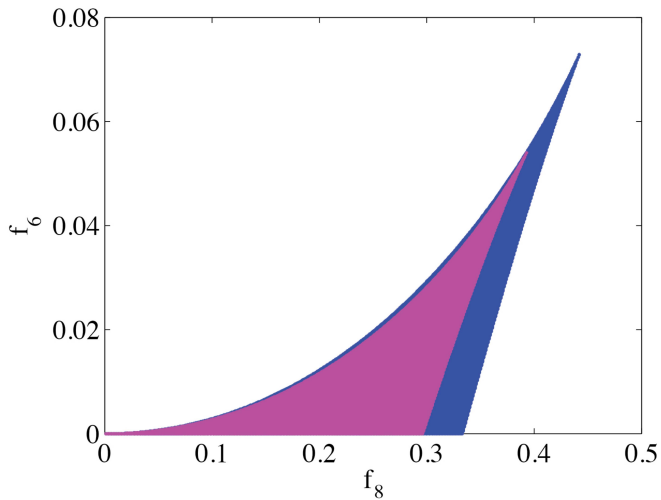}} 
\caption{(Color online) Multistability regions for HOKE models  in the $(f_6,f_8)$ parameter space. The inner (pink) area is described by Eq. (\ref{multistab_cond_topflats}) and 
corresponds to the existence of solitonic multistability (see Ref.~\cite{ultrasolitons}). The wider (purple) area corresponds to the existence of two branches of stable PWs, as described by Eq. (\ref{multistab_cond}).}
\label{multistability-regions}
\end{figure}


\subsection{Higher-Order Kerr nonlinearity}

HOKE nonlinearities have recently been introduced to describe the experimental observation of a saturation (and subsequent sign inversion) of the nonlinear correction to the
 refractive index at high optical intensities in gases~\cite{HOKEexp}. Among other implications, they have been theoretically argued to provide a new mechanism for the stabilization of optical filaments \cite{Cou1,MI5} without 
 the need for plasma-related effects~\cite{HOKEteor}. Such a nonlinear response can also give rise to the 
 existence of new localized light structures, called \emph{fermionic light} states~\cite{fermionic, ultrasolitons},  \emph{liquid light} states~\cite{michinel02,novoa09,fermionic, ultrasolitons}, or \emph{ultrasolitons}~\cite{ultrasolitons}. 
 By suitably rescaling $\Psi$ and the space-time coordinates, we will model a general nonlinear HOKE response as
\begin{equation}
\label{F_HOKE}
F(\vert \Psi\vert)=\sum_{q=1}^{n}(-1)^{q+1} f_{2q}\vert \Psi\vert^{2q},
\end{equation}
where, to be concrete, we will consider $f_{2q}>0$ for all $q$ and  $f_4=f_2=1$ like in the aforementioned CQ case. In the following discussion, motivated by the measurements in air and oxygen~\cite{HOKEexp,fermionic}, we will also assume that $n=4$, so that $f_{2q}=0$ for $q\ge 5$.

The dimensionless parameters entering the NLSE are related to the refractive index $n=n_0+\Delta n=n_0+ \sum_{q=1}^{4} n_{2 q} I^{q}$, where $n_0$ is the linear refractive index, by the relations
$\Delta n=({n_2^2}/{\vert n_ 4\vert })F$, $n_6=({n_4^2}/{n_ 2})f_6$ and $n_8=({n_4^3}/{n_ 2^2})f_8$, as shown in Ref. \cite{ultrasolitons}.
The only free parameters included in Eq. \ref{NLSE}  
are then $f_6$ and $f_8$,  which will be assumed to be positive taking into account the results of 
Ref. \cite{HOKEexp} for the optical response of common gases.

As we have chosen $A>0$, the stability condition Eq. \eqref{stability-cond}  can be written as $1-2V+3 f_6 V^2- 4 f_8 V^3<0$, where $V=A^2$. We can then have two different scenarios, depending on the roots of the algebraic equation

\begin{equation}
\label{multistab_eq}
1-2V+3 f_6 V^2- 4 f_8 V^3=0,
\end{equation}

namely: i) if Eq.  \eqref{multistab_eq} has only one positive real root $V_1$, the stability condition reads $A>\sqrt{V_1}$, while for $A<\sqrt{V_1}$ the PWs will undergo MI. ii) If Eq.  \eqref{multistab_eq}  has three positive real roots $V_1<V_2<V_3$, we find two different stability windows: $\sqrt{V_1}<A<\sqrt{V_2}$ and $A>\sqrt{V_3}$, whereas for $A<\sqrt{V_1}$ and for $\sqrt{V_2}<A<\sqrt{V_3}$ the corresponding PW will be modulationally unstable.
We can obtain an analytical condition on the parameters $f_6$ and $f_8$ for  Eq.  \eqref{multistab_eq} to have three real roots, which will delimitate the four stability and instability regions discussed above. This can be accomplished by calculating the ($f_6,f_8$) pairs for which the discriminant associated with the polynomial expression Eq. \eqref{multistab_eq} turns out to be negative, giving

 \begin{equation}
\label{multistab_cond}
27 f_6^3 - 9 f_6^2- 108 f_6 f_8+ 32 f_8 + 108 f_8^2<0.
\end{equation}


\begin{figure}[htbp]
\centerline{\includegraphics[width=0.75\columnwidth]{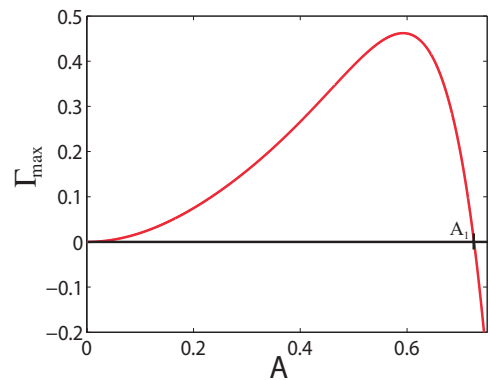}} 
\caption{(Color online) Representation of $\Gamma_{max}$ vs. $A$ for the HOKE model with $f_6=2.8$ and $f_8=3.9$. Whenever positive, $\Gamma_{\rm max}$ gives the exponential growth factor of the fastest growing filaments.}
\label{fig2-HOKE1A}
\end{figure}


As shown in Fig. \ref{multistability-regions}, this condition (purple area), allowing for the existence of two different parametric regions for stable PWs, is less stringent than the condition for the existence of multistable flat-topped soliton solutions reported in Ref.~\cite{ultrasolitons} (pink area), namely

\begin{equation}
\label{multistab_cond_topflats}
18225 f_6^3 - 5400 f_6^2- 77760 f_6 f_8+ 204802 f_8 +  93312 f_8^2<0.
\end{equation}


\begin{figure}[htbp]
\centerline{\includegraphics[width=0.75\columnwidth]{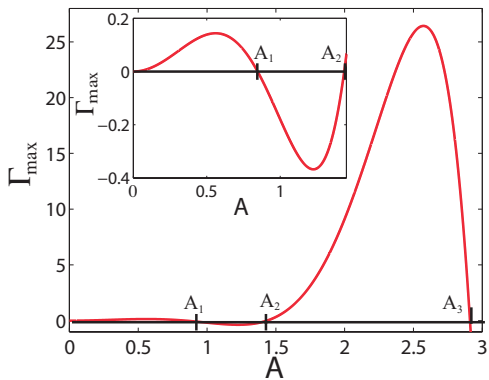}} 
\caption{(Color online) Representation of $\Gamma_{max}$ vs. $A$ for the HOKE model with $f_6=0.3$ and $f_8=0.02$. Whenever positive, $\Gamma_{\rm max}$ gives the exponential growth factor of the fastest growing filaments. Inset: close-up of the low-amplitude region.}
\label{fig2-HOKE1B}
\end{figure}


Hence, whenever two different soliton branches exist, thus implying soliton multistability as discussed in Ref.~\cite{ultrasolitons}, there will also be two different regions of stability for the PWs. However, the existence of two stability regions for the PWs does not necessarily imply the emergence of a second, \emph{ultrasolitonic} branch for the localized solutions.


\begin{figure}[htbp!]
\centerline{\includegraphics[width=1\columnwidth]{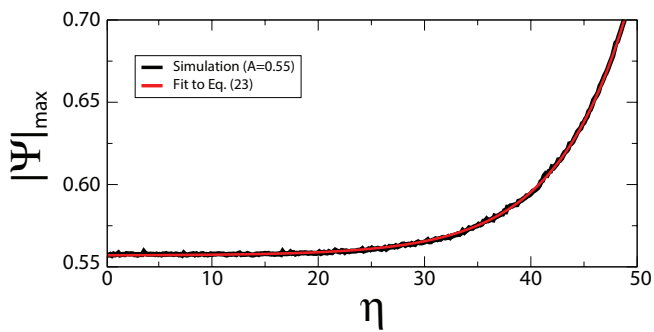}} 
\caption{(Color online) Evolution of the peak amplitude of a perturbed PW with A=0.55 in a HOKE medium ($f_6=0.3$, $f_8=0.02$). The results of the numerical simulation are represented by the black line, whereas the red line corresponds to a fit of the evolution to Eq. (\ref{exponential-growth}). The coefficients of the analytical fit are displayed in Table 2.}
\label{ajuste}
\end{figure}
Following the same procedure described above, for any value of $A$ belonging to one of the instability windows discussed above, we can compute the wavevector corresponding to the fastest growing perturbations as

\begin{equation}
  \label{kmax_HOKE} 
k_{\rm max}=\sqrt{\sum_{q=1}^{n}(-1)^{q+1} 2q f_{2q}A^{2q}},
\end{equation}
being the corresponding value of the maximum perturbation growth rate
\begin{equation}
  \label{Gammamax_HOKE} 
\Gamma_{\rm max}=\sum_{q=1}^{n}(-1)^{q+1} q f_{2q}A^{2q}.
\end{equation}


\begin{table}[htbp]
\begin{center}
\resizebox{1\columnwidth}{!}{
\begin{tabular}{c|c|c|c|c|c|c|c|c|}

\cline{2-8} & \multicolumn{5}{|c|}{Simulation} & \multicolumn{2}{|c|}{Theory}
\\
\hline
\multicolumn{1} {|c|} A & a ($\times10^{-5}$) & b & c & $R^2$ & $\Lambda_{sim}$ & $\Gamma_{max}$ &  $\Lambda$ 
\\
\hline
\multicolumn{1} {|c|} {0.30} & 2.784 & 0.090 & 0.304 & 0.9998 & 8.15$\pm$0.63 & 0.079 & 7.91 
\\
\hline
\multicolumn{1} {|c|} {0.60} & 29.704 & 0.210 & 0.607 & 0.9993  & 4.93$\pm$0.27 & 0.231 & 4.62
\\
\hline
\end{tabular}}
\caption{Results of the numerical simulations for the evolution of perturbed PWs with different amplitudes in the HOKE model with $f_6=2.8$, $f_8=3.9$. The first column gives the value of the initial amplitude $A$. The next columns give the values of $a$, $b$ and $c$ obtained from the fit of the evolution of the PW peak amplitude to an exponential growth, as described in Fig. \ref{ajuste}. $R^2$ gives the correlation of the fit. $\Lambda_{sim}$ is the numerical estimate for the initial width of the fastest emerging filaments. The last two columns reproduce the theoretical predictions of $\Gamma_{max}$ and $\Lambda=\pi/\sqrt{2\Gamma_{max}}$ for comparison. Notice that the parameter $b$ is the numerical counterpart of $\Gamma_{max}$.}
\end{center}
\label{table:HOKE1}
\end{table}

Let us illustrate these results with two concrete examples:

i) $f_6=2.8$ and $f_8=3.9$, corresponding to the central values of the $n_{2q}$ obtained in the experiment of Ref.~\cite{HOKEexp} to describe the propagation of ultrashort laser pulses in oxygen, namely 
$n_2=1.6 \times 10^{-19} cm^2/W$, $n_4=-5.2 \times 10^{-33} cm^4/W^2$, $n_6=4.8 \times 10^{-46} cm^6/W^3$, $n_8=-2.1 \times 10^{-59} cm^8/W^4$. In this case, Eq. \eqref{multistab_cond}  is not satisfied, so that Eq. \eqref{multistab_eq} has only a single real root, $V_1=0.526$, corresponding to the amplitude $A_1=\sqrt{V_1}=0.725$. The dependence of $\Gamma_{\rm max}$ on the PW amplitude $A$, as given by Eq. (\ref{Gammamax_HOKE}) is plotted in Fig. \ref{fig2-HOKE1A}. Notice that above the threshold $A>A_1$, $\Gamma_{\rm max}<0$ indicating that the corresponding PWs are linearly stable.

ii) $f_6=0.3$, $f_8=0.02$, corresponding to the parameters introduced in Ref.~\cite{ultrasolitons} to discuss a HOKE model predicting soliton multistability (Eq. \eqref{multistab_cond_topflats} is fulfilled in this case). Moreover, with these parameters Eq. \eqref{multistab_cond} is also satisfied, and so we get three real roots of Eq. \eqref{multistab_eq}, namely $V_1=0.716$, $V_2=2.060$ and $V_3=8.474$, whose square roots give the amplitudes $A_1=0.846$, $A_2=1.435$ and $A_3=2.911$, respectively. The corresponding dependence on $A$ of the values of $\Gamma_{\rm max}$, as  computed through Eq. \eqref{Gammamax_HOKE}, is plotted in Fig.  \ref{fig2-HOKE1B}. As we can clearly see in the figure, there are two stability windows, where $\Gamma_{\rm max}\le0$, for $A_1<A<A_2$ (see the inset of Fig. \ref{fig2-HOKE1B}) and for $A>A_3$. Conversely, there are two instability regions spanning the parametric ranges $A<A_1$ and $A_2<A<A_3$, respectively.

Interestingly, it has been recently reported that the HOKE response is usually non-instantaneous and it may also show a further effective dependence on both the intensity and pulse duration \cite{bejot13,kohler13}. Thus it is worth investigating different ranges of the ($f_6$,$f_8$) parameters, besides the values reported in~\cite{HOKEexp}. Our Fig. 1 can be used to infer how the effective parameters $f_6$ and $f_8$ vary without significantly modifying the qualitative behavior of the propagation of light in the media. Indeed, our analysis indicates that similar modulational instability landscapes are expected for values of the parameters $f_6$ and $f_8$ lying within the same region of Fig. 1. On the other hand, this may also open the possibility of the existence of multistability and ``ultrasoliton" states even in common optical media, provided that the effective ($f_6$,$f_8$) values enter in the multistability region for certain suitable intensity levels.
At this respect, \emph{ab-initio} and perturbative calculations for the hydrogen atom have revealed that an effective HOKE response can only be tolerated when all the higher-order corrections correspond to about $10\%$ of the cubic response~\cite{Spott14}. Under those constraints, we have checked that a sensitive HOKE response could be observed in a hypothetical medium featuring $f_6=0.3, f_8=0.02$ for intensities above $18~TW/cm^2$.

In the next section we will illustrate the theoretical results reported above by showing some numerical examples of the evolution of PWs in different nonlinear systems.


\section{Numerical Simulations}

Henceforth we will discuss the results of our numerical simulations for the evolution of the systems described in the previous section when different initial PWs are perturbed with low-amplitude white noise. We employed a standard split-step beam propagation method to solve Eq. \eqref{NLSE}~\cite{AgrawalKivshar}, being the step involving the linear part of the operator (diffraction) treated using a finite-differences scheme, the so-called Crank-Nicolson algorithm \cite{CNmethod,MuAd,Vudra12}, together with Neumann boundary conditions $\frac{\partial \Psi}{\partial r}=0$ at the borders. 
As this method makes efficient use of the whole computational domain for the physical results, we can mimic the evolution of PW-like fields featuring finite energy with an affordable computational effort. There are other approaches to simulate PW-like beams such as the consideration of large, but finite, flat-top beams restricting the observation plane to the central part of the domain, or the usage of periodic boundary conditions. These techniques, however, often introduce non-physical interactions in the propagation, which may eventually alter the results. Thus, in our case we can safely consider wide enough PWs and study the evolution of central areas of width much larger than the characteristic radii of the filaments born after the PWs destabilisation.


\begin{figure}[htbp!]
\centerline{\includegraphics[width=0.7\columnwidth]{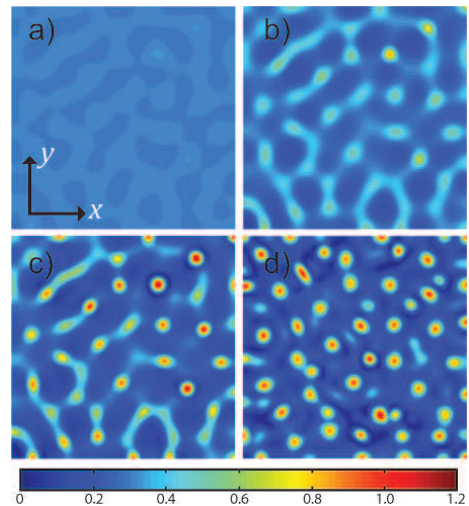}}
\caption{(Color online) Simulation of the evolution of a PW of initial amplitude $A=0.3$ (perturbed by white noise) in a system displaying HOKE nonlinearity with $f_6=0.3$ and $f_8=0.02$. The different panels show the two-dimensional amplitude distribution $\vert\Psi(x,y)\vert$ of the incoming field at different stages of its evolution (a) $\eta=80$, (b) $\eta=115$, (c) $\eta=125$ and (d) $\eta=150$. The spatial region displayed in the figures corresponds to $x,y\in [100,100]$.}
\label{hoke1a03}
\end{figure}


By monitoring the exponential growth of the maximum amplitude of the unstable fields, we can numerically obtain a direct estimation of the largest perturbation growth rates. To do so, we will fit the evolution of the maximum amplitude $\vert\Psi(\eta)\vert$ during the early stages of the PW destabilization to an exponential of the form

\begin{equation}
  \label{exponential-growth} 
\vert\Psi\vert= a e^{b\eta} + c,
\end{equation}
where the theoretical values for $b$ and $a+c$ correspond to $ \Gamma_{max}$ and $A\equiv\vert\Psi(0)\vert$, respectively. An example of this exponential behaviour of $\vert\Psi(\eta)\vert$ is shown in Fig. \ref{ajuste}, where we represent the early stages of evolution of a PW with initial amplitude A=0.55 in a HOKE nonlinear medium, together with a fit of the simulation results to Eq. (\ref{exponential-growth}). As it can be inferred from the graph, the agreement between simulation and theory is remarkable.

As the MI processes for the Kerr and CQ nonlinear models have already been extensively studied in the literature, we will concentrate the following discussion to the HOKE model for two different parametric regimes, namely ($f_6=2.8$, $f_8=3.9$) and ($f_6 = 0.3$, $f_8=0.02$).

\subsection{HOKE Model with $f_6= 2.8$ and $f_8= 3.9$.}

Let us consider a system displaying a HOKE nonlinearity of the form described in Eq. \eqref{F_HOKE} with $f_6 = 2.8$ and $f_8= 3.9$. In this case, as we have discussed above and as shown in Fig. \ref{fig2-HOKE1A}, the theory predicts the occurrence of MI for initial amplitudes $A <A_1=0.725$, and stability for $A>A_1$. The physical mechanism behind the stabilization of the PWs with amplitudes above the estimated threshold $A_1$ is the following: for small or moderate values of $A<A_1$, the positive terms entering the effective nonlinear refractive index correction $F$ (namely those proportional to $f_2$ and $f_6$) dominate, and hence $F$ increases with increasing $A$. This regime has been shown to support the existence of localized solitary waves satisfying an equation of state formally similar to that governing the dynamics of a degenerate Fermi gas~\cite{fermionic}. For values of $A>A_1$, however, the negative terms entering $F$ (namely those proportional to $f_4$ and $f_8$) dominate, and the function $F$ decreases with increasing $A$. Thus, in analogy with both Kerr and CQ systems, small perturbations on a PW background cannot be developed into hot filaments when the effective nonlinearity is self-defocusing, since light will tend to flow away from the regions of high intensity.

\begin{figure}[htbp!]
\centerline{\includegraphics[width=0.7\columnwidth]{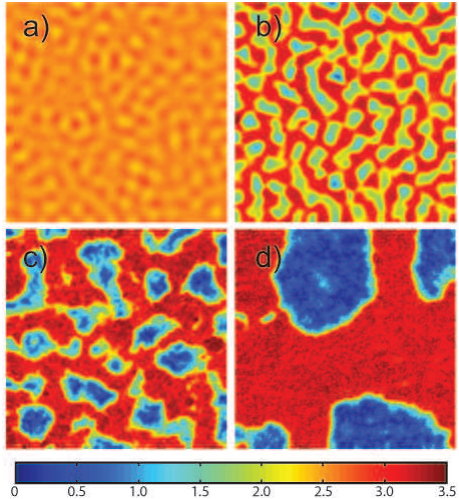}} 
\caption{(Color online) Simulation of the evolution of a PW of initial amplitude $A=2.55$ (perturbed by white noise) in a system displaying HOKE nonlinearity with $f_6=0.3$ and $f_8=0.02$. The different panels show the two-dimensional amplitude distribution $\vert\Psi(x,y)\vert$ of the incoming field at different stages of its evolution (a) $\eta=0.3$, (b) $\eta=0.5$, (c) $\eta=1$ and (d) $\eta=5$. The spatial region displayed in the figures corresponds to $x,y\in [10,10]$, since the characteristic size of the emerging filaments (see panel (a)) is much smaller than that of the filaments displayed in Fig. \ref{hoke1a03}.}
\label{hoke1a255}
\end{figure}

In this framework, the results of our simulations for the evolution of two perturbed PWs with amplitudes $A=0.3$ and $A=0.6$, i.e. well within the instability window, are summarised in Table 1. We have corroborated that the maximum amplitude evolves according to the exponential law given by Eq. (\ref{exponential-growth}) in a very good approximation. In particular, the numerical growth rates $b$ calculated from the fits to Eq. (\ref{exponential-growth}) agree with the theoretical predictions  $\Gamma_{max}$ obtained from Eq. \eqref{Gammamax_HOKE}. Furthermore, we have also compared the analytical estimations of the characteristic size of the arising filaments $\Lambda$ with the numerical results. To do so, we have monitored the size of the amplitude modulations $\Lambda_{sim}$ by subtracting the initial PW background at the very early stages of the PW destabilization. The results of this analysis are also included in Table 1, where the numerical value of $\Lambda_{sim}$ has been obtained by averaging over an ensemble of several filaments.


\subsection{HOKE Model with $f_6=0.3$ and $f_8=0.02$}

In this section we will analyze an example of HOKE model involving two instability windows. We choose $f_6=0.3, f_8=0.02$, which lies within the multistability region as shown in Fig. \ref{multistability-regions} above. In particular, we will study the dynamics of two PWs with initial amplitudes $A=0.3$ and $A=2.55$, belonging to the first and second instability regions introduced above, respectively (see section III-C and Fig. \ref{fig2-HOKE1B}). Figs. \ref{hoke1a03}-\ref{hoke1a255} summarize the main results of the simulations for the evolution of these two PWs. In Fig. \ref{hoke1a03}, we observe that the initially-perturbed PW featuring $A=0.3$ undergoes a redistribution of its energy as it evolves (panels (a)-(b)), giving rise to the formation of soliton-like localized structures (panel (c)). These solitary waves can then interact with each other in a complex way, provided that their phases are not correlated. Such dynamical behavior in nonlinear media has been first demonstrated in \cite{BE1,BE2}.

\begin{figure}[htbp!]
\centerline{\includegraphics[width=1\columnwidth]{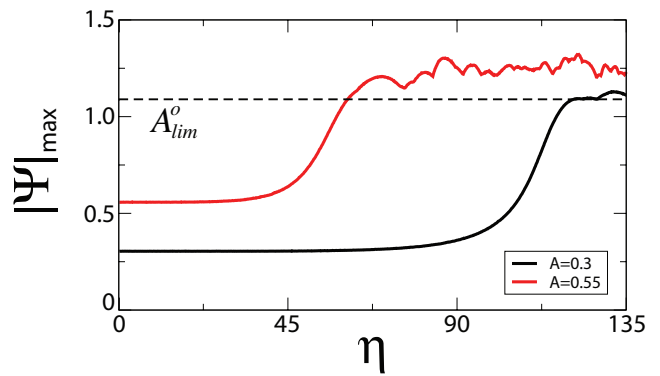}} 
\caption{(Color online) Evolution of the maximum amplitude of the PW fields with initial amplitudes $A=0.3,A=0.55$, belonging to the first instability window of the multistable HOKE model involving $f_6=0.3$ and $f_8=0.02$. Solid lines correspond to the numerical simulations, whilst the dashed line represents the limiting PW solution of the localized soliton branch of the system.}
\label{avstA}
\end{figure}

After a certain evolution period in which all emerging nonlinear beams exchange energy among them, some of the structures stabilize and form perturbed 2D solitons (panel (d)). Interestingly, all remnant solitons feature Gaussian-like shapes resembling those of the solitary waves found in Ref.~\cite{fermionic} for moderate amplitudes, i.e., far from the strong self-defocusing regime where flat-topped beams do exist~\cite{novoa09}. In particular, their amplitudes are slightly above the limiting value $A_{lim}^o=1.09$, which corresponds to the asymptotic PW solution of the \emph{ordinary}, CQ-like soliton branch~\cite{ultrasolitons}. This asymptotic behavior is reproduced in Fig.  \ref{avstA}. As we can appreciate in the figure, after a certain transient period the maximum amplitude of the field stabilizes, which turns out to coincide with the emergence of stable solitary waves in the system as a consequence of the MI-driven break up process.


\begin{table}[!hbt]
\begin{center}
\resizebox{1\columnwidth}{!}{
\begin{tabular}{c|c|c|c|c|c|c|c|c|}
\cline{2-8} & \multicolumn{5}{|c|}{Simulation} & \multicolumn{2}{|c|}{Theory}
\\
\hline
\multicolumn{1} {|c|} A & a ($\times10^{-5}$) & b & c & $R^2$ & $\Lambda_{sim}$ & $\Gamma_{max}$ &  $\Lambda$ 
\\
\hline
\multicolumn{1} {|c|} {0.30} & 5.388 & 0.085 & 0.304 & 0.9998 & 8.22$\pm$0.28 & 0.075 &  8.14
\\
\hline
\multicolumn{1} {|c|} {0.55} & 9.766 & 0.149 & 0.557 & 0.9998 & 6.02$\pm$0.23 & 0.144 &  5.86 
\\
\hline
\multicolumn{1} {|c|} {0.75} & 12.517 & 0.080 & 0.760 & 0.9974 & 8.20$\pm$0.81 & 0.082 &  7.76 
\\
\hline
\multicolumn{1} {|c|} {1.60} & 0.556 & 1.230 & 1.602 & 0.9997 & 2.26$\pm$0.24 & 1.116 &  2.10 
\\
\hline
\multicolumn{1} {|c|} {2.55} & 9.713 & 26.197 & 2.553 & 0.9999 & 0.53$\pm$0.05 & 26.361 &  0.43 
\\
\hline
\end{tabular}}
\caption{Results of the numerical simulations for the evolution of perturbed PWs with different amplitudes in the HOKE model with $f_6=0.3$, $f_8=0.02$. All parameters displayed are the same as those included in Table 1.}
\end{center}
\label{Table:HOKE2}
\end{table}



\begin{figure}[htbp!]
\centerline{\includegraphics[width=1\columnwidth]{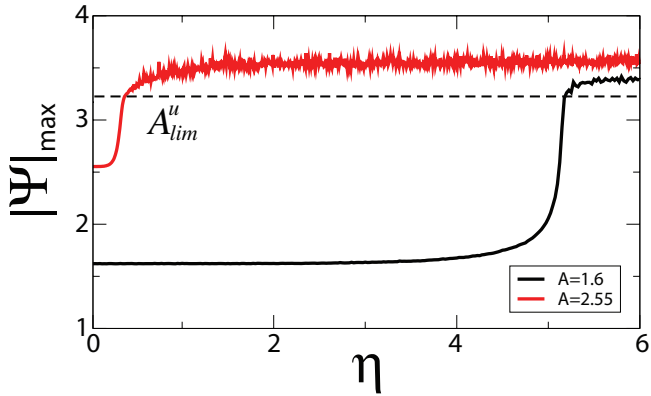}} 
\caption{(Color online) Same as in Fig. \ref{avstA} with initial amplitudes of the fields $A=1.6$ and $A=2.55$, belonging to the second instability window of the multistable HOKE model involving $f_6=0.3$ and $f_8=0.02$}
\label{avstB}
\end{figure}


On the other hand, Fig. \ref{hoke1a255} shows the dynamics of a perturbed PW featuring $A=2.55$, i. e. well within the second instability window displayed in Fig. \ref{fig2-HOKE1B}. Again, we observe a redistribution of the energy in the PW front with the subsequent formation of localized structures (panels (a)-(b)). However, in contrast to the behavior shown in Fig. \ref{hoke1a03}, the perturbed solitons emerging during the dynamics merge together to form large-scale structures with a nearly-homogeneous amplitude, as it can be inferred from the fields displayed in panels (c)-(d). This suggests that, for $A=2.55$, we can reach the threshold of existence of flat-topped states~\cite{michinel02, novoa09, CQsolitons}, so that the system evolves asymptotically towards the formation of a different PW-like structure, whose particular amplitude can be estimated analytically to be $A^u_{lim}=3.21$~\cite{ultrasolitons}. Again, the stabilization of the system at values slightly above $A^u_{lim}$ is corroborated by monitoring the dynamical evolution of the peak amplitude, as displayed in Fig. \ref{avstB}.

In addition, it is worth noticing that the overall dynamics displayed in Fig. \ref{hoke1a255} turns out to be much faster than that represented in Fig. \ref{hoke1a03}. The reason is that $\Gamma_{max}$ peaks close to $A=2.55$ according to the theory (see Fig. \ref{fig2-HOKE1B}), and its absolute value is more than two orders of magnitude higher than that corresponding to $A=0.3$, thus justifying the much shorter destabilization scale observed in Fig. \ref{avstB}. In order to reinforce this assertion, we have carried out the same numerical simulations described above with PWs featuring amplitudes $A=0.55$ and $A=1.6$, whose results are also depicted in Fig. \ref{avstA} and Fig. \ref{avstB}, respectively. We observe that, as expected, the destabilization occurs earlier since, in this range, $\Gamma_{max}$ increases for larger amplitudes following Eq. (\ref{Gammamax_HOKE}). In light of the same procedure described in Section IV-A above, we have also numerically computed $\Gamma_{max}$ by fitting the exponential growth of the PW amplitude at the early stages of MI to Eq. \eqref{exponential-growth}. The estimations of $\Gamma_{max}$, as well as the characteristic size of the fastest growing filaments $\Lambda_{sim}$, are summarized in Table 2 for different initial values of the PW amplitude. One can appreciate that the agreement between the analytical and numerical results is reasonable, even though there are small discrepancies inherent to the the fact that the first-order perturbation theory is only applicable when the amplitudes of the perturbations are very small.


\section{Conclusions.}

In this paper we have put forward a full analysis of the nonlinear process of modulational instability in systems described by nonlinear Schr\"odinger equations with arbitrary instantaneous nonlinear responses. In particular, the proposed theoretical approach allows for a complete description of the multiple regimes of stability and instability of plane waves in systems involving competing higher-order Kerr nonlinearities. All our analytical predictions for the stability domains, perturbation growth rates and characteristic filament sizes have been confirmed by direct numerical simulations of the evolution of unstable plane waves. This intriguing phenomenology could potentially be observed in a coherent atomic medium with a properly tailored nonlinear response. 

\acknowledgments

DT thanks the InterTech group of Valencia Politechnical University for hospitality during a research visit that was supported by the Salvador de Madariaga program of the Spanish Government. The work of DT is also supported by Xunta de Galicia through grant EM2013/002.


\end{document}